\documentclass{article}
\usepackage[final,nonatbib]{neurips_2024}
\usepackage[utf8]{inputenc} 
\usepackage[T1]{fontenc}    
\usepackage{hyperref}       
\usepackage{url}            
\usepackage{booktabs}       
\usepackage{nicefrac}       
\usepackage{microtype}      
\usepackage{xcolor}         
\usepackage{listings}%
\usepackage{colortbl}
\usepackage{makecell}
\usepackage{subfigure}
\usepackage{graphicx}%
\usepackage{multirow}%
\usepackage{amsmath,amssymb,amsfonts}%
\usepackage{amsthm}%
\usepackage{mathrsfs}%
\usepackage{textcomp}%

\newcommand{\aka}{\emph{a.k.a.},\ }
\newcommand{\et}{\emph{et al.}\ }

\title{Non-Overlapping Placement of Macro Cells based on Reinforcement Learning in Chip Design}

\author{
Tao Yu$^{1}$, Peng Gao$^{1}$, Fei Wang$^{2}$, Ru-Yue Yuan\\
$^1$Qufu Normal University \quad
$^2$Harbin Institute of Technology Shenzhen
}

\begin{document}

\maketitle

\begin{abstract}
Due to the increasing complexity of chip design, existing placement methods still have many shortcomings in dealing with macro cells coverage and optimization efficiency. Aiming at the problems of layout overlap, inferior performance, and low optimization efficiency in existing chip design methods, this paper proposes an end-to-end placement method, SRLPlacer, based on reinforcement learning. First, the placement problem is transformed into a Markov decision process by establishing the coupling relationship graph model between macro cells to learn the strategy for optimizing layouts. Secondly, the whole placement process is optimized after integrating the standard cell layout. By assessing on the public benchmark ISPD2005, the proposed SRLPlacer can effectively solve the overlap problem between macro cells while considering routing congestion and shortening the total wire length to ensure routability. Codes are available at \url{https://github.com/zhouyusd/SRLPlacer}.
\end{abstract}

\section{Introduction}\label{sec:1}

Placement in chip design represents the paramount and most labor-intensive phase\cite{r1,r2,tcasii1}. This critical step involves assigning locations within the chip layout to netlist components, encompassing macro and standard cells. Standard cells refer to elementary logic units, like logic gates, while macro cells denote more complex functional units, such as static random access memory (SRAM). The optimal layout enhances chip area utilization, timing performance, and routability. Consequently, the quest for methods to execute the macro cell placement task swiftly and efficiently has emerged as a prominent research topic in chip design.

Advanced automated placement methods play a crucial role in producing practical chip layouts, which markedly diminish the intricacy associated with chip designs and abbreviate the duration to market. Concurrently, optimizing layout can mitigate the circuit's resistance, inductance, and parasitic capacitance, thereby augmenting the signal transmission rate and diminishing the delay of logic gates\cite{r47,r48}. Moreover, this optimization can alleviate the leakage and thermal impacts, bolster reliability, decrease the occupied area, and amplify wafer utilization. These improvements are paramount for reducing manufacturing costs, underlining the significance of placement optimization in chip design.

Existing placement methods include partition-based, simulated annealing, and analytical placement, but these methods often show limitations when handling the placement of macro cells and standard cells. Traditional methods struggle to provide efficient solutions, especially when faced with complex interactions and high-dimensional action spaces. As an emerging technology, reinforcement learning (RL) can optimize different circuit performance metrics by designing various reward functions. The application of RL algorithms\cite{rl} has substantially increased the efficiency of chip design in recent years\cite{r19,filtering,r21}. These algorithms, notable for their enhanced adaptability and generalization capabilities, stand out as particularly effective for various chip design endeavors compared to conventional placement strategies. The ongoing advancement in computer hardware capabilities has also facilitated the broader adoption of end-to-end deep learning approaches\cite{r15,r16}. These sophisticated approaches can learn the intricate end-to-end mappings from unprocessed data, presenting more effective solutions to complex chip placement challenges. Nevertheless, such challenges often entail many constraints and objectives, encompassing wiring length, power consumption, and signal delay. Traditional placement methods focus on minimizing wiring length to optimize the chip layout. Yet, they exhibit limitations in addressing macro cell coverage effectively, revealing gaps in their practical application\cite{tcasii2,r47,r48}. Furthermore, considerations for subsequent processes, such as routing, are imperative, yet layouts generated by current placement methods suffer from suboptimal routability. This underscores the necessity for advancements in placement methods that can reconcile these multifaceted requirements and constraints.

This paper introduces a novel end-to-end placement method of macro cells in chip design, termed SRLPlacer, which leverages RL to tackle prevalent issues such as insufficient macro cell coverage, suboptimal macro cell layouts, and lackluster optimization efficacy observed in existing placement methods. Specifically, we utilize RL to transform the macro cell placement task into a Markov decision process\cite{markov}, enabling dynamic learning and optimizing placement strategies. Unlike traditional heuristic or deterministic methods, RL can easily transfer the experience accumulated during the macro cell placement process to new circuits, reducing the need for re-exploration and significantly decreasing the time required for optimization during placement. This transformation facilitates the learning and refining placement strategies pertinent to the macro cell\cite{r18}. Subsequently, SRLPlacer achieves comprehensive optimization throughout the placement process by incorporating the placed standard cells. A major innovation of this paper is the application of RL to the macro cell placement task, using a graph attention network (GAT)\cite{gat} in the policy network (\aka RL agent). GATs effectively model the intricate spatial and functional dependencies between macro cells, enhancing the network's generalization ability across different chip layouts and improving placement decision accuracy. This approach enables the RL framework to capture and utilize complex relationships that traditional methods cannot address. We also introduce a sophisticated dual reward strategy comprising immediate and overall rewards. Immediate rewards provide timely feedback during the placement process, helping mitigate the issue of sparse reward signals that can hinder learning efficiency. Overall rewards evaluate the final placement quality, considering key performance metrics such as total wire length and routing congestion. This dual reward strategy ensures local and global optimization objectives are met, a feature lacking traditional placement methods. Empirical evaluations conducted using standard benchmarks for placement demonstrate that the proposed SRLPlacer effectively addresses the macro cell coverage dilemmas and reduces the total wire length while maintaining adequate routability.

\section{Related Work}\label{sec:2}

The exploration and development of placement methods in chip design date back to the 1960s\cite{r5} and predominantly encompass three distinct categories: partition-based, simulated annealing, and analytical approaches.

Initially, the placement was approached using the divide-and-conquer strategy, hinging on partitioning methods. These approaches enjoyed widespread adoption and included pioneers such as the Kernighan-Lin\cite{r6} and Fiduccia-Mattheyses\cite{r7} algorithms. Drawing on graph theory's minimum cut principles, these approaches segmented the circuit into smaller blocks, refining the partitioning results through iterative enhancement. However, as circuit scales expanded, the simplistic nature of these partitioning approaches began to fall short of addressing the evolving practical requirements. This shortfall prompted the exploration of alternative placement approaches. Notably, the simulated annealing approaches\cite{r8} emerged as a prevalent choice for placement in chip design, characterized by their comprehensive optimization capability. These approaches employ random perturbations coupled with a strategy that tolerates less-than-ideal solutions, facilitating the surmounting of local optima in pursuit of the global optimum. Their robustness and adaptability have led to their widespread application, demonstrated by the development of simulated annealing-based algorithms like Hill-Climbing\cite{r9} and Tabu Search\cite{r10}, particularly in placement tasks. Despite the flexibility and comprehensive optimization capabilities of the simulated annealing approaches, their time-intensive nature poses challenges, particularly as the scale of chip designs continues to escalate.

Analytical approaches formulate mathematical models for placement tasks and derive optimal solutions. Typically, analytical approaches involve certain assumptions, like the differentiability of the objective function or treating placement cells as point-like entities. Among the analytical approaches, quadratic wire length\cite{r11} and nonlinear optimization\cite{r12} placement methods are notably prevalent. The quadratic wire length placement method, known for its computational expedience, nonetheless falls short in yielding optimal layouts compared to its counterpart. The nonlinear optimization placement method engages a more complex loss function to enhance the layout quality. Existing methods like ePlace\cite{r15} and RePlAce\cite{r16} have leveraged an electrostatics-inspired smooth density function and a nonlinear optimizer rooted in the Nesterov paradigm, showcasing superior performance in standard benchmarks. These methods conceptualize each netlist node as a positively charged particle, adjusting node positions based on inter-nodal repulsive forces, with the density function mirroring the system potential energy and facilitating node position updates through gradient optimization. To expedite computation, placement tasks may be segmented into subtasks and disseminated across multi-thread CPUs to efficiently manage the layouts of millions of standard cells. However, these methods necessitate substantial numerical computations, a domain where GPUs excel. Considering the analogies between placement optimization in the analytical approaches and neural network training, Lin \et proposed DREAMPlace\cite{r18}, adapting RePlAce\cite{r16} using the PyTorch toolkit\cite{r17} and harnessing GPU acceleration. As a result, DREAMPlace achieves a remarkable speed, operating over 30 times faster than RePlAce.

In the evolving landscape of artificial intelligence, adopting learning-based methodologies for placement in chip design, particularly those founded on RL\cite{rl}, has gained significant traction within academic and industrial communities. A noteworthy contribution from Google Research conceptualized the macro cell placement procedure as a series of sequential decisions, pioneering an end-to-end learning framework\cite{r19}. Within this framework, the RL agent (\aka policy network) finalizes the position of each macro cell incrementally, earning a specific reward with each placement action until the arrangement of the final macro cell is completed. The framework utilizes a graph neural network (GNN)\cite{gcn} within the value network to encapsulate the netlist, while the policy network employs a deconvolutional architecture to delineate the preferred place position for a macro cell. This innovative framework has dramatically accelerated the placement process, accomplishing in hours what traditional placement approaches might take weeks or months to complete.
Further, Vashisht \et have merged RL with heuristic strategies to address the placement problem, introducing a cyclical framework that intertwines RL with simulated annealing\cite{r20}. The RL agent is tasked with adjusting the spatial relationships between netlist components, while the simulated annealing algorithm expands on this preliminary arrangement to explore the layout solution further. Cheng \et have broken new ground by proposing DeepPR\cite{r21}, an end-to-end learning method for joint placement and routing macro cells. DeepPR deploys two distinct RL agents: one for placement and another for routing.

Despite advancements in chip design, the persistent escalation in circuit integration and complexity continues to render the placement task exceedingly formidable. The design of a chip profoundly influences its performance and is intricately connected to critical aspects such as power consumption, layout, and reliability\cite{r2}. An optimal placement method can significantly mitigate signal transmission delays, curtail physical area, decrease power utilization, and bolster reliability and resilience against interference.

\section{Methodology}\label{sec:3}

\begin{figure}[t!]
    \begin{center}
        \includegraphics[width=\linewidth]{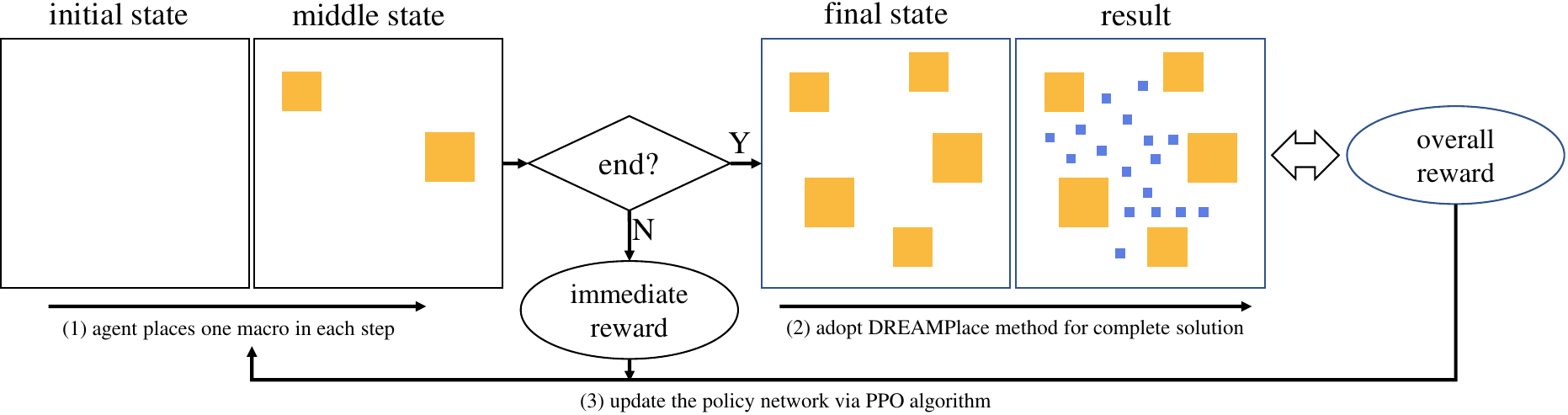}
    \end{center}
    \caption{Overview pipeline of our proposed placement method, SRLPlacer. Orange squares indicate macro cells, while blue dots represent standard cells.}
    \label{fig:1}
\end{figure}

In this section, we describe the problem and design constraints associated with the placement process in chip design. The placement problem is as follows: Given a set of macro cells with specific dimensions and pin configurations, determine their locations on the chip layout to ensure that no two macro cells overlap. Additionally, the layout must account for the positions of standard cells and optimize key design constraints, including total wire length, routing congestion, and macro cell layout density. The primary goal of the placement process is to achieve non-overlapping placement of all macro and standard cells while optimizing various circuit design constraints. The total wire length affects the transmission delay of the circuit; shorter total wire length results in lower transmission delay, better circuit performance, and reduced chip size. Routing congestion indicates the difficulty of the subsequent routing process and the heat dissipation performance of the circuit; lower routing congestion simplifies the routing process and enhances heat dissipation. Lower macro cell layout density allows more standard cells to be distributed around the macro cells, reducing the total wire length and avoiding macro cell coverage issues. This paper leverages an RL agent (\aka policy network) to determine the optimal positions for each macro cell sequentially. Once a macro cell's position is determined, it checks if all macro cells have been placed. If so, their positions are fixed, and all standard cells are placed using the gradient optimization-based placement method, DREAMPlace\cite{r18}. During the placement of standard cells, the half-perimeter wire length (HPWL) of a randomly sampled state is used as a reward to encourage the RL agent to explore more states. After obtaining the final placement solution, the total wire length, layout density, and routing congestion are used as the overall reward for update the policy network via the PPO algorithm\cite{ppo}. If the macro cell placement is not yet finished, the layout density and wire length of the current macro cell placement state are used as the immediate reward to update the RL agent. The overall pipeline of our proposed method is illustrated in Figure \ref{fig:1}. Using RL to solve a new problem often requires implementing an environment for interaction with the agent and designing the critical elements of the Markov decision process. For the placement problem faced in this paper, the environment corresponds to the chip layout quality assessment process, the state corresponds to whether the layout currently marks placed macro cells, the action corresponds to selecting an unmarked position in the layout that can place the current macro cell, and the reward corresponds to indicators such as HPWL, layout density, and routing congestion.

\subsection{State Space}\label{sec:3-1}

This paper designs a two-dimensional discrete 0-1 state space to mark the positions where macro cells can be placed in the chip layout. The state representation comprises four parts: netlist metadata, netlist graph $\mathcal{H}$, macro feature $\mathcal{F}$, and current macro $\mathrm{id}$. $\mathcal{H}$ describes the adjacency relationships among all macro cells before outlined in the netlist metadata. $\mathcal{F}$ describes the geometric information of all macro cells, including size, number of pins, and types of pins. $\mathrm{id}$ represents the identity number of the macro cell to be placed currently. In the specific implementation, a $N\times N$ matrix represents $\mathcal{H}$, where $N$ indicating the number of macro cells; a vector represents the netlist metadata; another $N\times 4$ matrix represents $\mathcal{F}$; $\mathrm{id}$ is an integer, used as an index to extract mask information processed by the GNN\cite{gcn}, for the policy network's output layer to decide the position where the macro cell corresponding to the current macro id should be placed.

\subsection{Action Space}\label{sec:3-2}

The core design concept of the action space is to provide a highly flexible and precise framework to support complex macro cell placement decisions. In chip design, the placed position of macro cells directly impacts critical metrics such as performance, power, and area. Therefore, the action space needs to be able to finely represent all possibilities of macro cell placed position, allowing the method to explore and find the optimal placement strategy. Thus, after constructing a state space that reflects the features and connection information of macro cells, this paper also carefully designs a matching action space.
The action space includes $W\times W$ actions, represented by a $W\times W$ matrix, where $W$ is the size of the chip layout. The matrix covers all possible macro cell placement positions on the chip layout. With such representation, the action space can finely describe the placement strategy of macro cells with $W\times W$ actions, each corresponding to a possible placed position for a macro cell. During the placement process, the action space is not just a candidate set for macro cell placement decisions, and it can also reflect the dynamic changes of chip layouts. As each macro cell is successfully placed, the action space will be updated to show the new placement state and the remaining placement possibilities. This dynamic updating mechanism allows the method to adapt to changes during placement, continually optimizing the placement strategy for the remaining macro cells.

\begin{figure}[t!]
    \begin{center}
        \includegraphics[width=\linewidth]{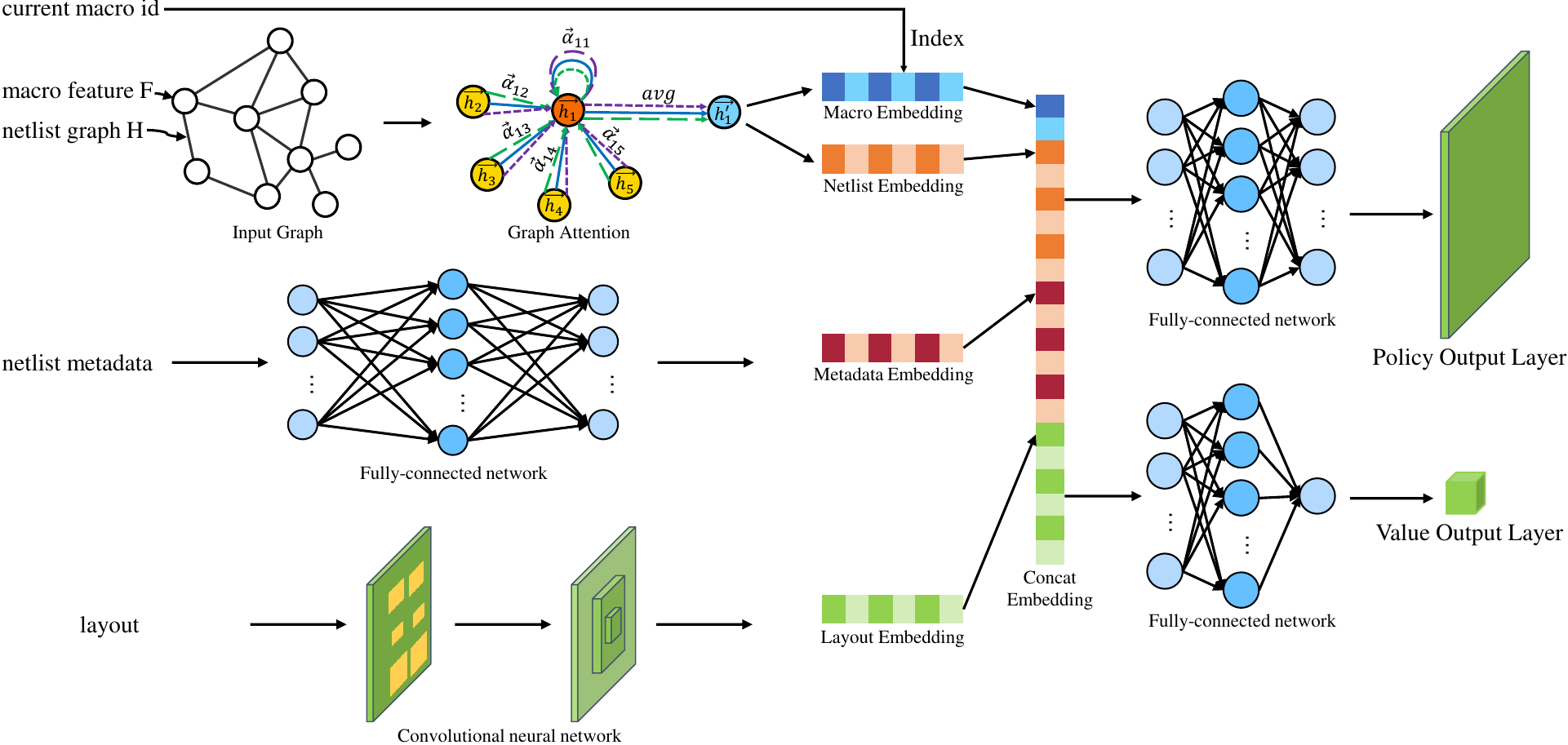}
    \end{center}
    \caption{Overview architecture of the policy network.}
    \label{fig:2}
\end{figure}

\subsection{Rewards}\label{sec:3-3}

In the placement process, an effective reward design is crucial because it directly affects the learning efficiency of the RL agent and the quality of the final layout. The reward design proposed in this paper carefully considers the critical performance indicators of chip layout, namely HPWL, layout density, and routing congestion (see Section \ref{sec:4-2}), to ensure that the resulting layout meets physical requirements and electrical performance. The basic idea of the reward design is to guide the policy network in making the best macro cell placement decisions by precisely calculating the impact of each action on layout quality. The reward of SRLPlacer consists of two parts: one is the overall reward $R_t$ after the placement of all macro and standard cells has been completed, and the other is the immediate reward $R_i$ after one macro cell is placed. Together, these two rewards constitute the feedback for learning the policy network, aiming to achieve an optimal layout.

The basic idea of the reward design is to guide the policy network in making the best macro cell placement decisions by precisely calculating the impact of each action on layout quality. The reward of SRLPlacer consists of two parts: one is the overall reward $R_t$ after the placement of all macro and standard cells has been completed, and the other is the immediate reward $R_i$ after one macro cell is placed. Together, these two rewards constitute the feedback for learning the policy network, aiming to achieve an optimal layout.

\subsubsection{Overall Reward}\label{sec:3-3-1}

Although chip placement aims to minimize power and area, traditional methods require several hours to complete an evaluation, making this approach unfeasible for an RL agent that needs continuous iterative learning. Optimizing wire length can reduce transmission delays to some extent; optimizing the routing congestion can reduce the difficulty of subsequent routing. In this way, the performance and routability of macro cells placement can be optimized more effectively. Therefore, we propose a overall reward $R_t$, including wire length and routing congestion, to maximize performance and routability simultaneously as follows,
\begin{equation}\label{eq:1}
  R_t=-\alpha_t\cdot WL(P,H)-\beta_t\cdot C(P,H)
\end{equation}
where $\alpha_t$ and $\beta_t$ are hyperparameters that respectively represent the importance of total wire length and routing congestion; $P$ represents the state when the positions of all macro and standard cells have been placed; $H$ represents the connection information between macro and standard cells; $C$ is used to measure the level of routing congestion for a given state; $WL$ is used to calculate HPWL of all connections appearing in the netlist for placed units. For each net connection $n_i$, with its endpoint set $\{x_i\}$ and $\{y_i\}$, HPWL can be calculated as follows,
\begin{equation}\label{eq:2}
  HPWL(n_i)=(\max\{x_i\}-\min\{x_i\})+(\max\{y_i\}-\min\{y_i\})
\end{equation}

HPWL can be approximated as the connection length on the critical path. Optimizing this metric can help reduce signal transmission delay and jitter and decrease the chip size. The level of routing congestion (\aka $C$) can be approximated as the complexity of subsequent routing. Optimizing this metric ensures that the routing stage can be completed smoothly, meets the timing requirements, and avoids crosstalk issues between multiple wires, ensuring signal integrity.

\subsubsection{Immediate Reward}\label{sec:3-3-2}

Because the RL agent cannot receive reward signals long before the standard cell placement by DREAMPlace is completed, the initial macro cell placement process becomes a random RL task. During training, the inability to obtain rewards for a long time can lead to low-quality layouts. Considering that the number of macro cells is much less than that of standard cells, the wire length of already placed macro cells is insufficient to determine the quality of the final layout. Still, the area occupied by macro cells is much greater than that of standard cells. The layout density of macro cells can better represent the quality of the current layout. We propose a immediate reward $R_i$ that includes macro cell placement wire length and layout density to encourage the RL agent to explore potential high-quality layouts as follows,
\begin{equation}\label{eq:1}
  R_i=-\alpha_i\cdot WL(\tilde{P},H)-\beta_i\cdot D(\tilde{P})
\end{equation}
where $\alpha_i$ and $\beta_i$ are hyperparameters to balance the importance of immediate reward wire length and layout density; $WL$ and $H$ are as described in Section \ref{sec:3-3-1}; $\tilde{P}$ represents the state of placed macro cells; $D$ is used to approximate the layout density for a given placement state.

SRLPlacer primarily determines the positions of macro cells sequentially, then identifies the placement of standard cells by DREAMPlace, jointly optimizing the layout problem. The main decision-making occurs during the macro cell placement phase. The introduction of immediate rewards is to prevent the RL agent from experiencing slow convergence due to long periods without reward signals. Macro cells are often hundreds or thousands of times larger than standard cells. To avoid overlapping macro cells and leave adequate space for standard cells, we introduced the macro cell layout density (\aka $D$) into the immediate reward. This metric reflects the dispersion of macro cells on the layout. Since macro cells are sufficiently dispersed and large, the HPWL can effectively reflect the connection length of the critical path during the macro cell placement process. Optimizing this metric can also improve circuit performance and decrease chip size. Additionally, this metric helps balance the dispersion, preventing excessively dispersed macro cell layouts that could lead to larger chip sizes.

\subsection{Policy Network}\label{sec:3-4}

The policy network consists of multiple modules, each containing several submodules, which process the input data and ultimately output a probability distribution corresponding to the actions, as shown in Figure \ref{fig:2}. We use the policy network to learn how to place macro cells to maximize the rewards. Unlike traditional neural networks, our policy network is unsupervised learning. Its most important feature is that it needs to interact with the environment to obtain rewards. In the macro cell placement problem, the policy network must output a probability distribution corresponding to each action based on the current state and action space. An action is selected based on this probability distribution and applied to the environment, providing a new state and reward. The policy network aims to learn an optimal strategy through interaction with the environment, such that the optimal action under the current state yields the maximum rewards.

The input layer of the policy network needs to be capable of receiving and processing two-dimensional feature information from the state space. The policy network backbone adopts a multi-layer GAT architecture\cite{gat} to identify critical factors affecting placement quality, such as the relative positions of macro cells and the spatial distribution within the layout. The output layer maps the high-dimensional features extracted from the backbone onto a vector with dimensions equal to the size of the action space, where each element represents the probability of placing a macro cell. To achieve this mapping, the output layer uses the Softmax function as the activation to ensure that the sum of all action probabilities equals 1 and that each action probability is non-negative. Thus, the policy network can choose an action to execute randomly or greedily based on the output probability distribution. Specifically, to better map the output results of the policy network to the positions on the layout, this paper reshapes the output dimensions of the policy network into two dimensions.

\subsection{Environment}\label{sec:3-5}

The environment needs to be capable of updating the placement state based on the actions of the RL agent. This includes calculating the positions of newly placed macro cells, updating the features of the placed area, such as HPWL, layout density, and routing congestion, and detecting potential constraint conflicts, such as macro cell overlaps. We divide the $W\times W$ action space (\aka layouts) into 20 equal grids both horizontally and vertically, resulting in a total of 400. The estimated number of routing lines in each grid is weighted summed (with weights decreasing uniformly from 1 at the centroid of the layouts to 0 at the edges) to approximate the level of routing congestion. The layout density is determined by the average distance between each macro cell and its nearest neighboring macro cell. HPWL is calculated for each pair of connected cells, and the overall HPWL of the circuit is the summation of the HPWLs of all connections. When the RL agent selects an illegal action, it is given a negative reward to guide it toward learning placement strategies that meet constraints. If the updated placement strategy is valid, calculate the immediate reward of the current layout and use the square of the immediate reward of the current layout minus the immediate reward of the layout before performing the action as the score for this action of the RL agent. If the immediate reward after the action is less than before, the difference is negative; otherwise, it is positive. Additionally, the environment also controls the start and end of training rounds. When a training round begins, the environment is initialized. Specifically, the placement state is reset to zero, i.e., all the placed cells are removed. As the RL agent optimizes parameters, the environment determines whether all macro cells have been placed. If all macro cells have been placed, standard cells will be placed using DREAMPlace. After that, the total reward is calculated, serving as the overall reward and ending the current training round. The storage module is then called to save important training data from this iteration, such as states and rewards, which can be used for further performance optimization and analysis.

\section{Experiments}\label{sec:4}

\subsection{Benchmark and Settings}

The validation experiments for SRLPlacer were conducted on the ISPD2005 benchmark\cite{ispd2005}. This dataset was first introduced at the 2005 International Symposium on Physical Design (ISPD) and is mainly used for evaluating and comparing the performance of routing methods. ISPD2005 consists of different test cases, each representing a routing problem that must be solved. These test cases simulate the challenges in the chip design, including numerous netlists and complex routing areas, requiring the evaluated methods to effectively handle large-scale routing problems while optimizing critical paths to reduce total wire length and congestion. Although the ISPD2005 dataset was initially designed for routing tasks, its inclusion of high-density routing areas and complex constraints makes it a valuable resource for evaluating placement algorithms. Using the ISPD2005 dataset for placement experiments allows focusing on exploring how to effectively place macro and standard cells and how to deal with the impact of routing constraints on placement strategies. The statistics of the right test cases in ISPD2005 are shown in Table \ref{tab:1}.

The proposed method is implemented in Python with Pytorch on a server with one Intel$^\circledR$ Core$^\texttt{TM}$ i9-10900X CPU @ 3.70GHz CPU with 128GB RAM, and one NVIDIA$^\circledR$ Geforce$^\circledR$ RTX 3080TI GPU with 12GB VRAM. The PPO algorithm\cite{ppo} is used for policy network learning, and the DREAMPlace method\cite{r18} is adopted for arranging standard cells. During training, totally $32$ epochs are performed with the Adam optimizer\cite{adam} and $2.5\times 10^{-4}$ learning rate.

\begin{table}[t!]
\centering
\caption{Statistics of ISPD-2005 benchmark dataset.}
\label{tab:1}
\begin{tabular}{ccccccc}
\toprule
Test Cases & Total Cells (K) & Nets (K) & Movable Macros & Fixed Macros & Design Density    \\
\midrule
adaptec1 & 211 & 221 & 514 & 29 & 75.71\% \\
adaptec2 & 255 & 266 & 542 & 24 & 78.56\% \\
adaptec3 & 451 & 466 & 710 & 13 & 74.53\% \\
adaptec4 & 496 & 516 & 1309 & 20 & 62.67\% \\
bigblue1 & 278 & 284 & 551 & 9 & 54.19\% \\
bigblue2 & 558 & 577 & 948 & 22136 & 61.80\% \\
bigblue3 & 1097 & 1123 & 1227 & 66 & 85.65\% \\
bigblue4 & 2177 & 2230 & 659 & 7511 & 65.30\% \\
\bottomrule
\end{tabular}%
\end{table}

\begin{table}[t!]
\centering
\caption{The results of the ablation study (in HPWL $\times10^6$).}
\label{tab:2}
\begin{tabular}{ccccc}
\toprule
Test Cases & SRLPlacer$_{GAT}$ & SRLPlacer$_{R_i}$ & SRLPlacer$_{GCN}$ & SRLPlacer \\
\midrule
adaptec1 &  91.8731	& 85.1221	& 96.7781	& \textbf{81.5813} \\
adaptec2 & 129.9153	&126.1026	&136.9510	&\textbf{121.8029} \\
adaptec3 & 257.9472	&254.1146	&268.8911	&\textbf{247.2351} \\
adaptec4 & 263.1078	&255.8651	&268.1386	&\textbf{245.1513} \\
bigblue1 & 169.1003	&165.9527	&175.0495	&\textbf{159.7567} \\
bigblue2 & 236.1222	&228.6257	&243.9162	&\textbf{213.0494} \\
bigblue3 & 469.0379	&467.8111	&476.2681	&\textbf{453.4381} \\
bigblue4 & 959.7115	&962.0723	&992.8007	&\textbf{941.8300} \\
\bottomrule
\end{tabular}%
\end{table}

\begin{table}[t!]
\centering
\caption{The results of the quantitative analysis (in HPWL $\times10^6$).}
\label{tab:3}
\resizebox{\columnwidth}{!}{%
\begin{tabular}{cccccc}
\toprule
Test Cases & fs50\cite{r47} & K\&D\cite{r48} & Google\cite{r19} & DREAMPlace\cite{r18} & SRLPlacer (Ours) \\
\midrule
adaptec1 &  101.1700&	127.8900&	86.5757&	 83.7214&	 \textbf{81.5813} \\
adaptec2 & 122.8900&	157.6500&	124.1251&	 congestion check failure&	\textbf{121.8029} \\
adaptec3 & 287.7200&	293.2800&	257.9568&	 256.9236&	\textbf{247.2351} \\
adaptec4 & 337.2200&	352.0100&	255.7707&	 congestion check failure&	\textbf{245.1513} \\
bigblue1 & 174.5700&	163.4400&	168.4568&	 congestion check failure&	\textbf{159.7567} \\
bigblue2 & 285.4300&	322.2200&	142.0375&	 congestion check failure&	\textbf{133.0494} \\
bigblue3 & 471.1500&	656.1900&	460.5816&	 congestion check failure&	\textbf{453.4381} \\
bigblue4 & 1040.0500&	1403.7900&	1039.9760&	 congestion check failure&	\textbf{941.8300} \\
\bottomrule
\end{tabular}%
}
\end{table}

\begin{table}[t!]
\centering
\caption{Comparison of efficiency of different placement methods.}
\label{tab:4}
\begin{tabular}{ccccc}
\toprule
Test Cases & RePlAce\cite{r16} & SRLPlacer (Ours) \\
\midrule
adaptec1 & 45.8s&	\textbf{16.6}s \\
adaptec2 & 75.0s&	\textbf{16.0}s \\
adaptec3 & 187.5s&	\textbf{37.5}s \\
adaptec4 & 354.0s&	\textbf{41.6}s \\
bigblue1 & 83.3s&	\textbf{29.0}s \\
bigblue2 & 195.8s&	\textbf{45.8}s \\
bigblue3 & 320.8s&	\textbf{79.0}s \\
bigblue4 & 925.0s&	\textbf{154.0}s \\
\bottomrule
\end{tabular}%
\end{table}

\begin{figure}[t!]
\begin{center}
    \subfigure[adaptec1]
    {\label{fig:11a1}\fbox{\includegraphics[width=0.435\linewidth]{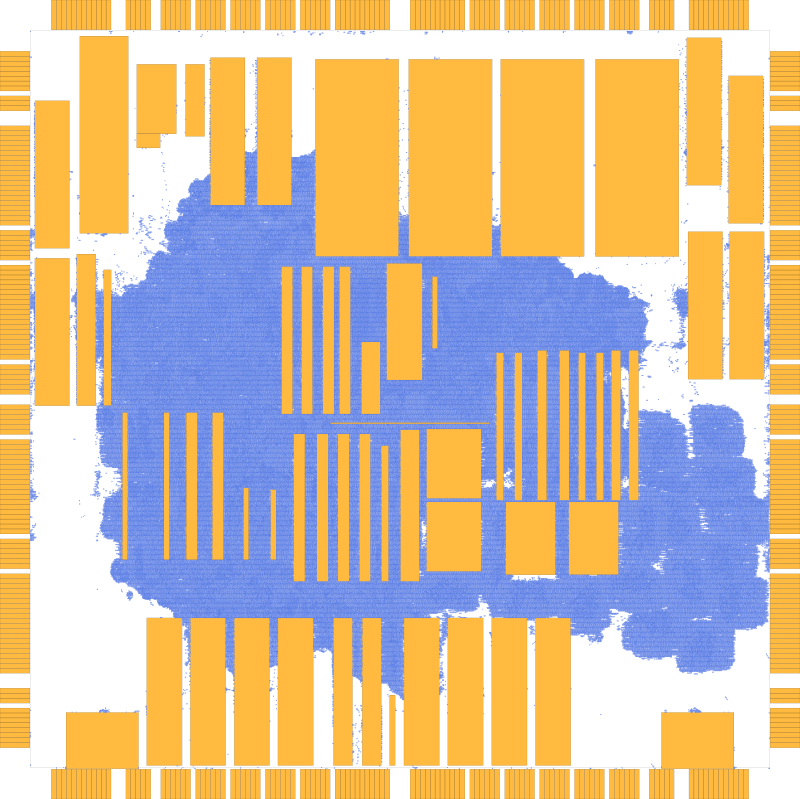}}}\hspace{0.2em}
    \subfigure[adaptec2]
    {\label{fig:11b1}\fbox{\includegraphics[width=0.435\linewidth]{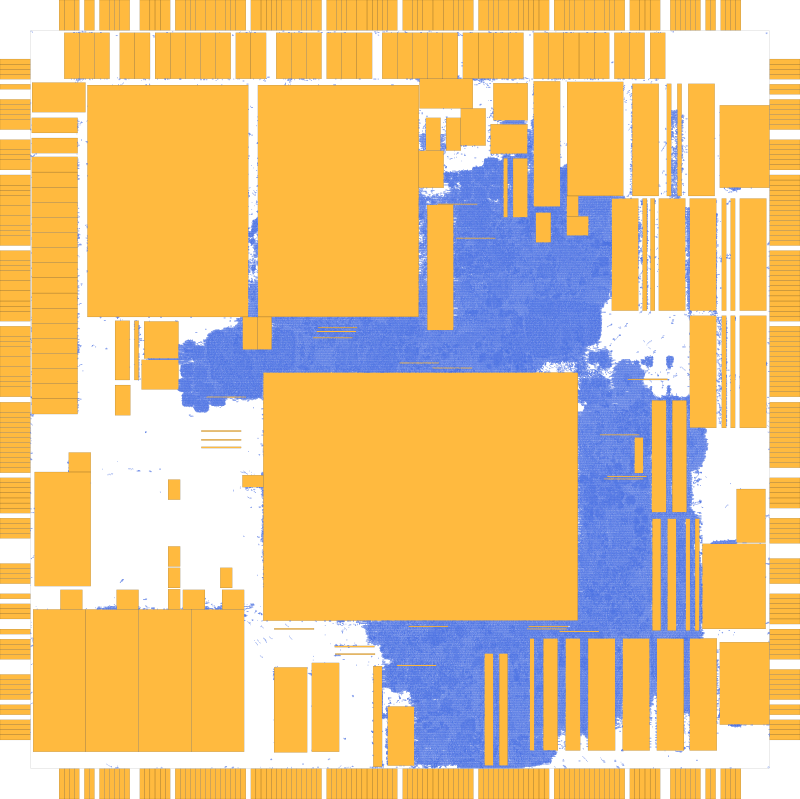}}}
    \vfill
    \subfigure[adaptec3]
    {\label{fig:11c1}\fbox{\includegraphics[width=0.435\linewidth]{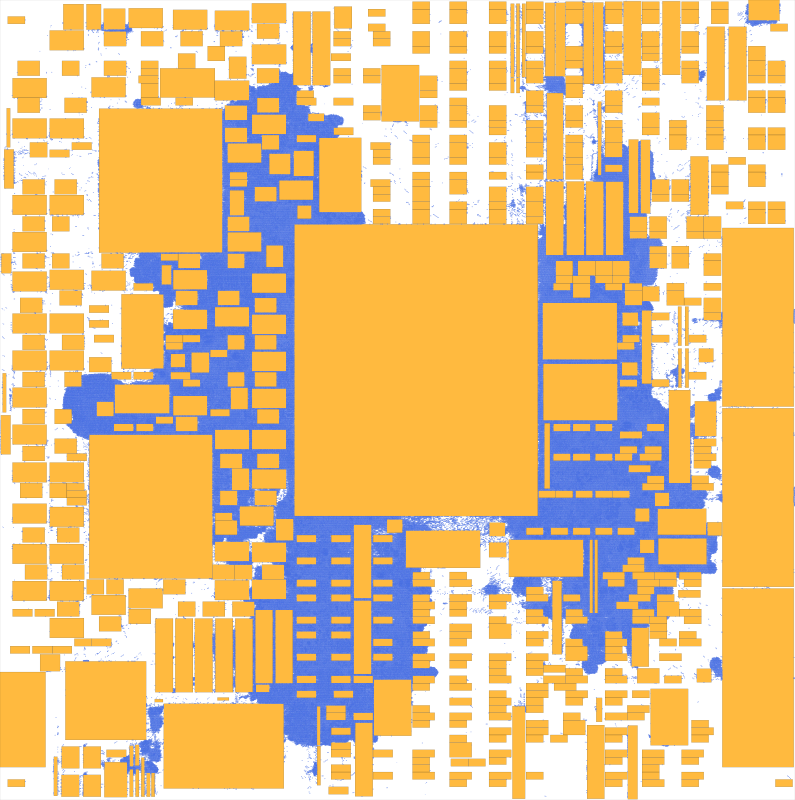}}}\hspace{0.2em}
    \subfigure[adaptec4]
    {\label{fig:11d1}\fbox{\includegraphics[width=0.435\linewidth]{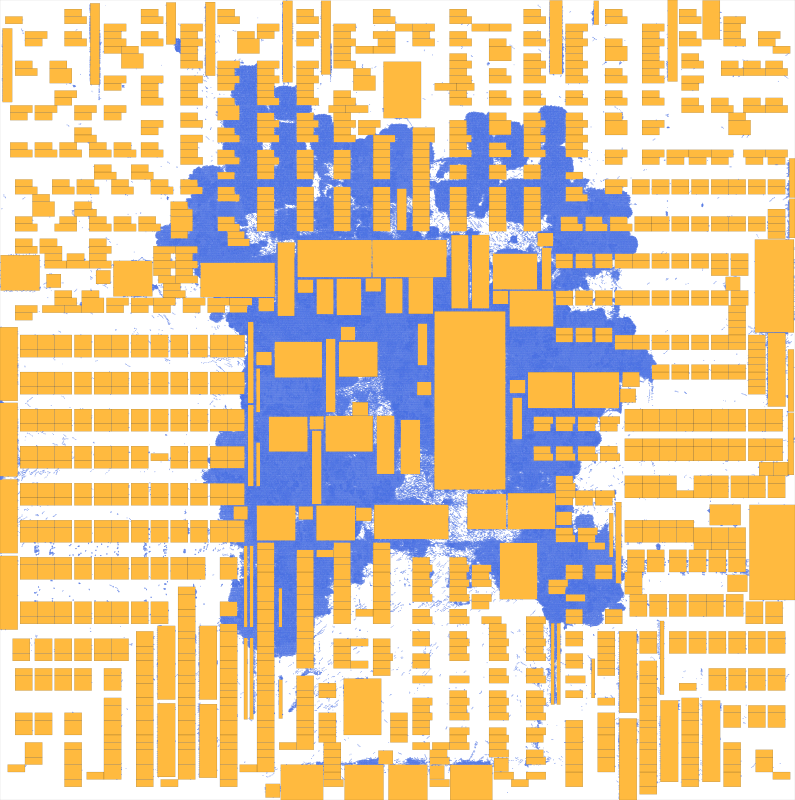}}}
\end{center}
\caption{Visualized placement results of four adaptec-class test cases of our SRLPlace in the ISPD2005 benchmark. The orange square represents the macro cell, and the blue region represents the layout of standard cells.}
\label{fig:11}
\end{figure}

\begin{figure}[t!]
\begin{center}
    \subfigure[bigblue1]
    {\label{fig:11a2}\fbox{\includegraphics[width=0.435\linewidth]{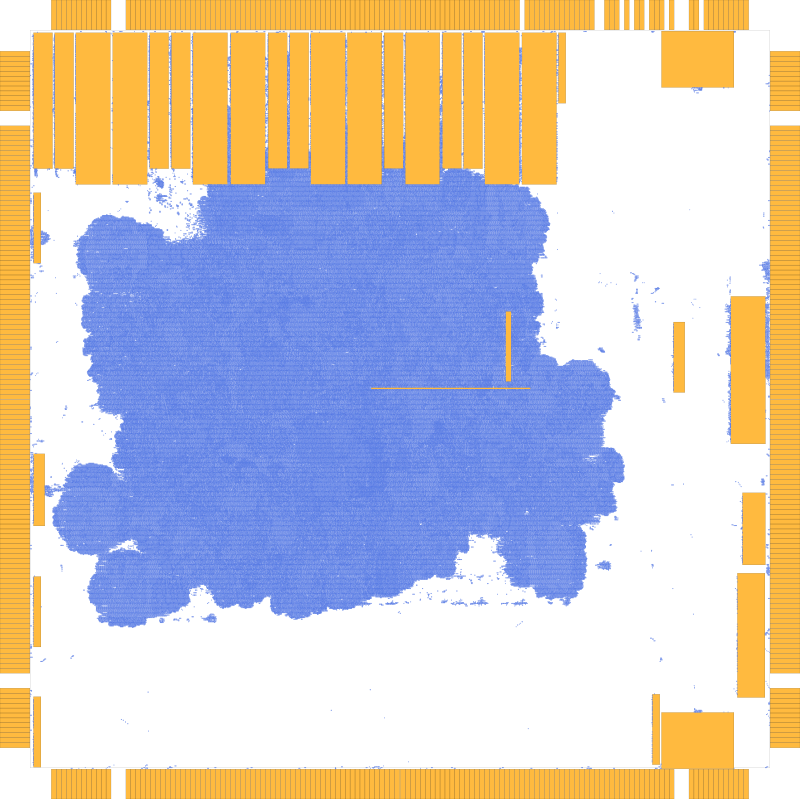}}}\hspace{0.2em}
    \subfigure[bigblue2]
    {\label{fig:11b2}\fbox{\includegraphics[width=0.435\linewidth]{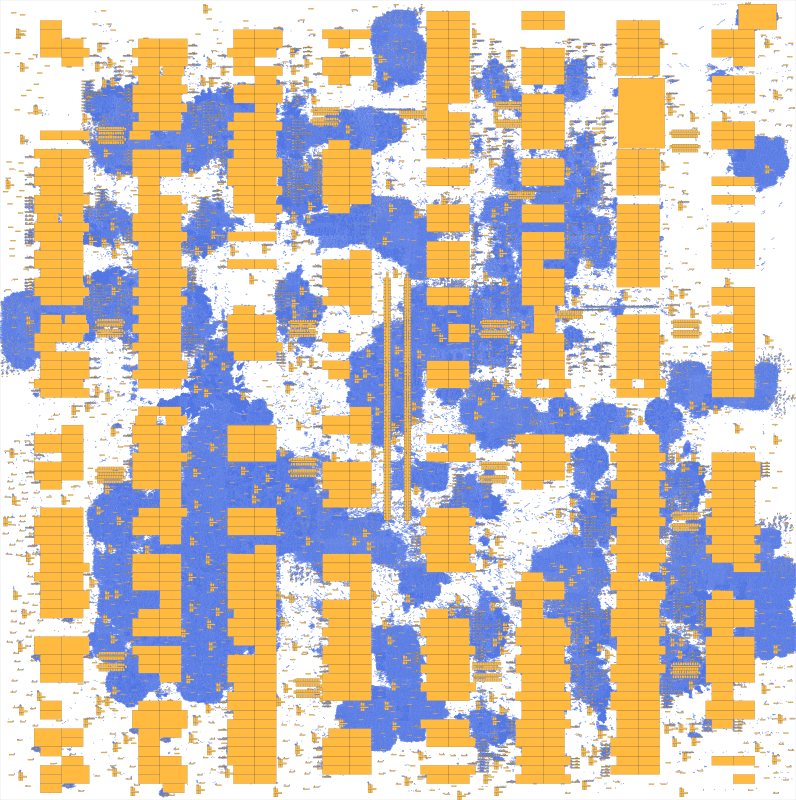}}}
    \vfill
    \subfigure[bigblue3]
    {\label{fig:11c2}\fbox{\includegraphics[width=0.435\linewidth]{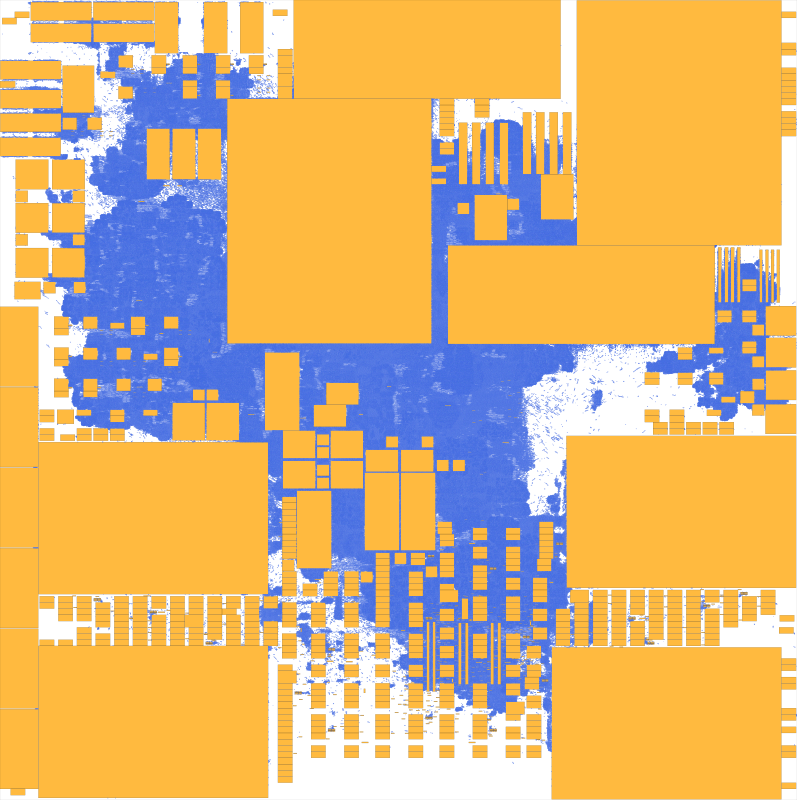}}}\hspace{0.2em}
    \subfigure[bigblue4]
    {\label{fig:11d2}\fbox{\includegraphics[width=0.435\linewidth]{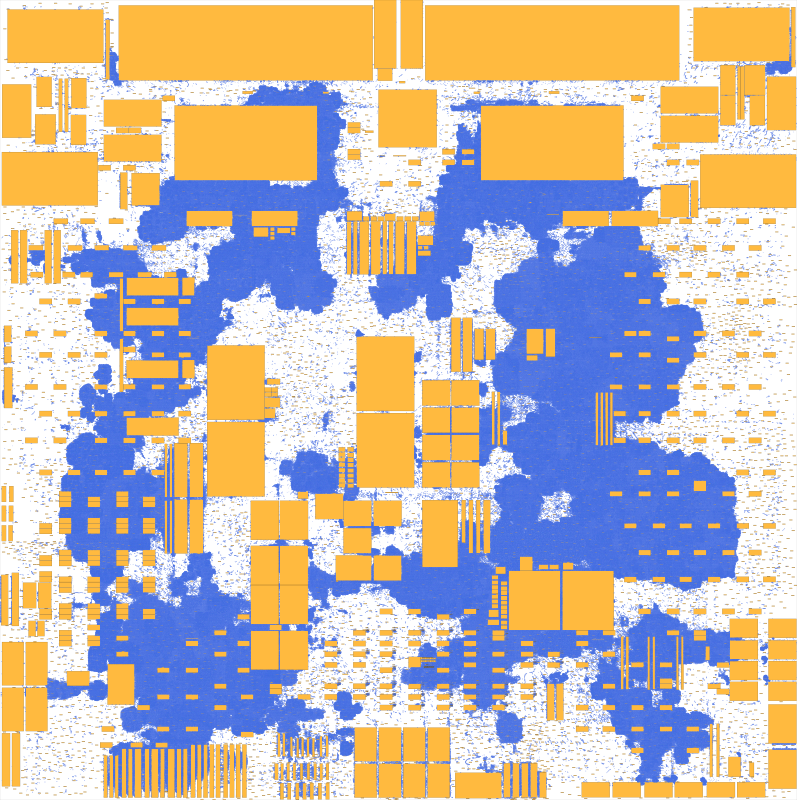}}}
\end{center}
\caption{Visualized placement results of four bigblue-class test cases of our SRLPlace in the ISPD2005 benchmark. The orange square represents the macro cell, and the blue region represents the layout of standard cells.}
\label{fig:12}
\end{figure}

\subsection{Metrics}\label{sec:4-2}

Commonly, the following metrics are used to evaluate the performance of placement methods.

\emph{HPWL} is a widely used metric to estimate the wiring length of a layout. It calculates the half-perimeter length of the smallest bounding box enclosing all pins of a netlist. Lower HPWL values indicate better layout since they typically correspond shorter to interconnect lengths and potentially lower signal transmission delays.

\emph{Layout density} refers to the ratio of the occupied area to the total available area within the layout. Higher density can indicate efficient use of available area, but excessive density can lead to heat dissipation problems and manufacturing challenges.

\emph{Routing congestion} measures the crowdedness of routing paths in the layout. High congestion can lead to routing problems and affect the chip's manufacturability and performance. Congestion can be evaluated by the percentage of routing paths used in the most congested area or the average usage across the entire chip.

\subsection{Ablation Study}\label{sec:4-3}

To validate the effectiveness of GAT and immediate rewards, this paper conducted a series of ablation studies to assess different combinations of whether the policy network backbone uses GCN or GAT and whether immediate rewards are introduced. Specifically, the research is divided into the following parts: using only GAT for placement (SRLPlacer$_{GAT}$), using only the immediate rewards (SRLPlacer$_{R_i}$); and using GCN without immediate rewards (SRLPlacer$_{GCN}$). The results of the ablation study are shown in Table \ref{tab:2}. It can be seen that using GAT and introducing immediate rewards both lead to a significant reduction in the HPWL for all 8 test cases. This indicates that compared to GCN, GAT can more effectively capture and utilize the relationships between nodes when dealing with placement problems, thus achieving a better layout. Introducing immediate rewards provides more timely feedback to the RL agent, helping it learn better placement strategies more quickly. From the results of these ablation experiments, replacing GCN with GAT in the policy network backbone and combining it with immediate rewards significantly enhances the placement performance of SRLPlacer. This further demonstrates the effectiveness and importance of GAT and immediate rewards in the problem of macros placement.

\subsection{Quantitative Analysis}\label{sec:4-3}

In exploring the application of RL methods in macros placement tasks, SRLPlacer has demonstrated significant performance advantages, especially in handling macro cell placement problems. Compared with other popular placement methods, SRLPlacer successfully generated effective placement strategies and achieved the best performance in the critical performance metrics of HPWL across eight different scale and complexity test cases, as shown in Table \ref{tab:3}. This achievement can be attributed to SRLPlacer's deep understanding and innovative solutions to both macro cell and standard cell placement problems, especially its ability to effectively avoid macro cell overlaps and optimize congestion and density, a significant challenge in traditional placement methods.

The experimental design meticulously considered the characteristics of different test cases, such as the ratio of macro cells to standard cells and the size of the chip area, ensuring the comprehensiveness and reliability of the experimental results. The sequential learning technique adopted by SRLPlacer is crucial for accurately predicting and optimizing the placement positions of macro cells, which was fully demonstrated in the comparative experiments. Compared with layout tools like DREAMPlace, SRLPlacer showed its absolute advantages in handling complex chip layouts with a large number of macro cells, mainly due to its efficient algorithm design, which significantly improves execution efficiency while maintaining layout quality.

The performance of SRLPlacer is reflected in its excellent result of HPWL and the significant improvement in placement efficiency, as shown in Table \ref{tab:4}. This indicates that SRLPlacer is not only suitable for current chip design needs but also has the potential to meet future VLSI design challenges. Nevertheless, the application and development of SRLPlacer still face a series of challenges, including enhancing its adaptability to new chip designs and integrating more advanced optimization techniques. Future research directions could consider these challenges to promote the continuous optimization of SRLPlacer.

\subsection{Visualized Layout}\label{sec:4-4}

Figures \ref{fig:11} and \ref{fig:12} respectively shows the visualized results of SRLPlacer on eight test cases from ISPD2005: adaptec1, adaptec2, adaptec3, adaptec4, bigblue1, bigblue2, bigblue3, and bigblue4.
These visualized results show that all placed macro cells are non-overlapping, indicating that SRLPlacer can effectively handle the distribution problems of macro cells. Furthermore, the results show that the interconnections between different cells are pretty reasonable, with reduced crossings and overlaps, which is crucial in lowering signal transmission delays and interference.

\section{Conclusion}\label{sec:5}

The SRLPlacer method proposed in this paper adopts a phased strategy that can simultaneously solve both macro and standard cell placement problems. SRLPlacer models and learns the interactions between macro cells, converting the placement problem into a Markov decision process, which can generate high-quality placement strategies. The routing congestion is also considered to ensure the routability of the final placement strategy while shortening the wire length. In experiments on the public benchmark ISPD2005, the performance and efficiency of SRLPlacer have shown significant improvements over several advanced existing placement methods. The results indicate that SRLPlacer is an innovative solution capable of effectively solving the non-overlapping macro cell placement problem in chip design. Future work aims to improve the parallelization and hardware acceleration capabilities of SRLPlacer to accommodate VLSI design tasks. As an end-to-end method, SRLPlacer also needs to be expanded to adapt to other design objectives relative to placement tasks, such as power optimization, thermal characteristic optimization, and signal integrity enhancement.

\bibliographystyle{num}
\bibliography{bibliography}

\end{document}